\documentstyle[preprint,aps,epsfig]{revtex}

	{\arrayclosep 0.14em\begin{eqnarray}}{\end{eqnarray}}
\newfont{\Fr}{eufm10}   %%% German characters %%%
\def\be{\begin{equation}}
\def\ee{\end{equation}}
\def\bea{\begin{eqnarray}}
\def\eea{\end{eqnarray}}
\def\b{\begin{eqnarray*}}
\def\e{\end{eqnarray*}}
\def \no{\nonumber}
\def \( {\left(}
\def \) {\right)}
\def \la {\langle}
\def \ra {\rangle}
\def\[{\left[}
\def\]{\right]}
\def\lsim {~^{<~}_{\sim~}}
\def\gsim {~^{>~}_{\sim~}}
\def \d     {\partial}
\def \f     {\frac}
\def \tr    {\makebox{tr}}

\def\Journal#1#2#3#4{{#1}{\bf #2}, #4 (#3)}
\def\BOOK#1#2{{\it #1} {(#2)}}
\def\MEET#1#2#3#4#5{{#1} {\it #2}, {#3}, edited by {#4} {(#5)}}
\def\MEETtmp#1#2#3#4{{#1} {\it #2}, {#3}, {(#4)}}

\def\NPB{{ Nucl. Phys.} \bf B}
\def\NPBPS{{ Nucl. Phys.} B (Proc. Suppl.) }
\def\PLB{{ Phys. Lett.}  B }

\def\PRD{{ Phys. Rev.} D }
\def\PRC{{ Phys. Rep.} \bf C}

\def\PTPS{{ Prog. Theor. Phys. Suppl. }}

\def \Gmumu   {G_{\mu\mu}}

\def \tilGmunu{\tilde{G}_{\mu\nu}}
\def \Amu     {A_{\mu}}
\def \Amua    {A_{\mu}^a}

\def \Aalmu   {A^\alpha_{\mu}}
\def \Umu     {U_{\mu}}
\def \Mmu     {M_{\mu}}
\def \umu     {u_{\mu}}
\def \kmu     {k_{\mu}}
\def \knu     {k_{\nu}}

\def \ftau3i2 {\f{\tau^3}{2}}
\def \ftaui2  {\f{\tau^a}{2}}

\def \Meff    {M_{\rm off}}
\def \RchA    {R_{\rm off}}
\def \vH    {\vec{H}}
\def \vAmu  {\vec{A}_\mu}

\def \hDdmu {\hat{D}_{\mu}}

\begin{document}
\draft
\title{
Off-diagonal Gluon Mass Generation and 
Infrared Abelian Dominance in the Maximally Abelian Gauge 
in Lattice QCD} 
\author{Kazuhisa Amemiya and Hideo Suganuma}
\address{Research Center for Nuclear Physics (RCNP), Osaka University\\
Mihogaoka 10-1, Ibaraki, Osaka 567-0047, Japan \\
E-mail: amemiya@rcnp.osaka-u.ac.jp} 
%%%%%%%%%%%%%%%%%%%%%%%%%%%%%%%%%%%%%%%%%%%%%%%%%%%%%%%%%%%%%%
% abstract 
%%%%%%%%%%%%%%%%%%%%%%%%%%%%%%%%%%%%%%%%%%%%%%%%%%%%%%%%%%%%%%
\maketitle
\begin{abstract}
We study effective mass generation of off-diagonal gluons 
and infrared abelian dominance 
in the maximally abelian (MA) gauge. 
Using the SU(2) lattice QCD, 
we investigate the propagator and the effective mass 
of the gluon field in the MA gauge 
with the U(1)$_3$ Landau gauge fixing. 
The Monte Carlo simulation is performed 
on the $12^3 \times 24$ lattice with $2.2 \le \beta \le 2.4$, 
and also on the $16^4$ and $20^4$ lattices with 
$2.3 \le \beta \le 2.4$. 
In the MA gauge, 
the diagonal gluon component $A_\mu^3$ shows long-range propagation, 
and infrared abelian dominance is found 
for the gluon propagator. 
In the MA gauge, 
the off-diagonal gluon component $A_\mu^\pm$ behaves as 
a massive vector boson with the effective mass 
$M_{\rm off} \simeq 1.2$ GeV in the region of $r \gsim 0.2$ fm, 
and its propagation is limited within short range. 
We conjecture that infrared abelian dominance can be interpreted as 
infrared inactivity of the off-diagonal gluon due to 
its large mass generation induced by the MA gauge fixing. 

\end{abstract}
\vspace{0.5cm}
\pacs{PACS number(s):12.38.Gc, 12.38.Aw, 11.15.Ha}

\vfill\eject
%%%%%%%%%%%%%%%%%%%%%%%%%%%
%%%%%%%%introduction%%%%%%%
\section{Introduction}
\label{section:intro}
The quark-confinement mechanism is one of the most important subjects 
in the nonperturbative QCD (NP-QCD) and in the hadron physics. 
From a phenomenological point of view, 
quark confinement is characterized by 
the linear inter-quark potential $V(r) \sim \sigma r$ 
with the hadronic string tension 
$\sigma \simeq 1$GeV/fm, which is obtained from the 
Regge trajectories \cite{GrRev} of hadrons 
and also from the lattice QCD simulation \cite{Rothe}.
This confining force arises from the one-dimensional squeezing 
of the gluonic flux between the quark and the anti-quark, 
which is actually observed in the lattice QCD 
simulation \cite{Hay}. 

On the quark-confinement mechanism, 
Nambu \cite{YN} first proposed 
the dual-superconductor picture in 1974 
using the concept of the electromagnetic duality \cite{Drc} 
in the Maxwell equation 
and the analogy with the one-dimensional squeezing of the 
magnetic flux as the Abrikosov vortex in the type-II 
superconductor \cite{Abr}. 
In the dual-superconductor picture, 
color-magnetic monopoles are assumed to condense, 
and then the color-electric flux between 
the quark and the anti-quark is squeezed as a one-dimensional tube 
due to the dual Higgs mechanism 
\cite{YN,GtH75,SM}.

The possibility of the appearance of monopoles in QCD 
was pointed out by 't~Hooft in 1981 
using the concept of the abelian gauge fixing 
\cite{GtH81}, which is defined by the diagonalization of 
a gauge-dependent variable 
as $\Phi [A_\mu(x)] \in {\rm su}(N_c)$. 
In fact, the abelian gauge fixing is a partial gauge fixing which 
remains the abelian gauge degrees of freedom on 
the maximally torus subgroup $U(1)^{N_c-1} \subset SU(N_c)$. 
As a remarkable feature in the abelian gauge, 
color-magnetic monopoles appear as the topological objects relating to 
the nontrivial homotopy group, 
$\Pi_2(SU(N_c)/U(1)^{N_c-1})=Z_\infty^{N_c-1}$ 
%\cite{GtH81,SST,ichiemp}. 
[9-12].
 
Then, the quark-confinement mechanism can be physically interpreted 
with the dual Meissner effect in the abelian dual Higgs theory 
[10,11,13-15] 
under the two assumptions of abelian dominance 
\cite{GtH81,SST,TS88and89,EI} 
and monopole condensation [17-19] 
in the abelian gauge. 
Here, abelian dominance firstly named by Ezawa and Iwazaki 
in 1982 \cite{EI} means that only the diagonal gluon component 
plays the dominant role for the NP-QCD phenomena like confinement. 
In 1990's, the approximate relations of 
abelian dominance for quark confinement 
[18,20-22] 
and chiral symmetry breaking \cite{OM,Wlf} 
are numerically shown in the lattice QCD Monte-Carlo simulation 
in the maximally abelian (MA) gauge [18,20-26],  
which is a special abelian gauge in case of $N_c=2$. 

Abelian dominance is usually discussed on the role of 
the diagonal (abelian) component of the gluon field. 
However, in terms of the off-diagonal gluon, 
abelian dominance can be expressed that off-diagonal gluon components 
are inactive at the infrared scale of QCD 
and can be neglected for the argument of NP-QCD. 
Here, abelian dominance as infrared inactivity of off-diagonal gluons 
may be interpreted with a large off-diagonal gluon mass. 
In this paper, we study the gluon propagator 
and the effective mass of off-diagonal gluons in the MA gauge, 
using the SU(2) lattice QCD Monte-Carlo simulations.

\section{Infrared Abelian Dominance and Mass-Generation Hypothesis 
on Off-diagonal Gluons in the MA Gauge} 

In the maximally abelian (MA) gauge, 
the nonabelian gauge symmetry ${\rm SU}(N_c)$ of QCD 
is reduced into the abelian gauge symmetry ${\rm U(1)}^{N_c-1}$, and 
accordingly diagonal gluons can be distinct from off-diagonal gluons. 
For instance, 
diagonal gluon components behave as neutral gauge fields like photons, 
and off-diagonal gluon components behave as charged matter fields 
on the residual abelian gauge symmetry 
\cite{GtH81,SST}.

In the MA gauge, 
abelian dominance for quark confinement and chiral-symmetry breaking 
has been shown as the approximate relation 
in the lattice QCD Monte-Carlo simulations 
[18,20-24], 
and only diagonal gluons seem to be 
significant to the infrared QCD physics, 
which we call ``infrared abelian dominance". 
In this section, we consider infrared abelian dominance and
the off-diagonal gluon property in the MA gauge in the SU(2) QCD. 
In terms of the off-diagonal gluon, 
infrared abelian dominance means that off-diagonal gluon components 
are inactive and their contribution can be neglected 
at the infrared scale of QCD in the MA gauge. 

As a possible physical interpretation for infrared abelian dominance, 
we conjecture that the effective mass of the off-diagonal gluon 
$\Amu^\pm \equiv \f{1}{\sqrt 2}( \Amu^1 \pm i \Amu^2)$ 
is induced in the MA gauge. 
If $\Amu^\pm$ acquires a large effective mass $\Meff$, 
the off-diagonal gluon propagation is limited within the short range 
as $r \lsim \Meff^{-1}$, 
since the massive particle is propagated within the inverse of its mass. 
Hence, in the mass-generation hypothesis of $A_\mu^\pm$ 
in the MA gauge, the massive off-diagonal gluons cannot 
mediate the long-range interaction 
and their direct contribution is expected to be negligible 
at the infrared scale, 
although there remains the indirect contribution of 
off-diagonal gluons through the coupling to the diagonal gluon. 
(This is similar to the negligible contribution of heavy mesons 
for the nuclear force at a large distance, 
where only light mesons as pions play the significant role.) 

The mass-generation hypothesis on off-diagonal gluons 
in the MA gauge is formally expressed in QCD as follows. 
The QCD partition functional in the MA gauge is expressed as 
\bea
Z^{\rm MA}_{\rm QCD}
	&\equiv&\int{\cal D}\Amu \exp \Bigl\{ iS_{\rm QCD}[\Amu] \Bigr\}
				 \delta(\Phi_{\rm MA}^{\pm}[\Amu])
				 \Delta_{\rm FP}[\Amu],
\eea
where $\Delta_{\rm FP}[\Amu]$ denotes 
the Faddeev-Popov determinant \cite{cheng}.
Here, $\Phi_{\rm MA}^{\pm}[\Amu]$ 
denotes the off-diagonal component of 
$
\Phi_{\rm MA}[\Amu]
	\equiv
	[\hDdmu,[\hat D^\mu,\tau^3 ]],
$ which is diagonalized in the MA gauge \cite{ichiemp,Sgnmykis,ichiead}, 
and therefore the MA gauge fixing is provided by 
$\delta(\Phi_{\rm MA}^{\pm}[\Amu])$. 
The mass-generation hypothesis of off-diagonal 
gluons $\Amu^{\pm}$ is expressed as 
\bea
Z^{\rm MA}_{\rm QCD}
	&=&\int{\cal D}\Amu^3  
			\exp \left\{ iS_{\rm Abel}[\Amu^3] \right\}
		\int{\cal D}\Amu^{\pm} 
			\exp \left\{ iS_{\rm off}^{M}[\Amu^\pm]\right\}
		{\cal F} [\Amu].
							\label{eqn:4simc01}
\eea
Here, $S_{\rm off}^{M}[\Amu^\pm]$ denotes 
the U(1)$_3$-invariant action of the off-diagonal gluon 
with the effective mass $M_{\rm off}$ as 
\bea
S_{\rm off}^{M}[\Amu^\pm] &\equiv& 
	\int d^4 x 
	\left\{ -\f{1}{2}
	   \( D^{\rm Abel}_\mu A_\nu^+ - D^{\rm Abel}_\nu A_\mu^+ \) 
 	   \( D_{\rm Abel}^{\mu*} A^\nu_- - D_{\rm Abel}^{\nu*} A^\mu_- \)
	   + \Meff^2 \Amu^+A^\mu_-
	 \right\}.\no \\
							\label{eqn:4simc02}
\eea
with the U(1)$_3$ covariant derivative  
$D^{\rm Abel}_\mu \equiv \d_\mu +ieA^3_\mu$. 
Here, $S_{\rm Abel}[\Amu^3]$ is the effective action 
of the diagonal gluon component, 
and 
${\cal F}[\Amu]$ is a U(1)$_3$-invariant smooth functional 
in comparison with 
$\exp \left\{ iS_{\rm off}^{M}[\Amu^\pm] \right\}$ at least 
in the infrared region. 

Since the massive off-diagonal gluons cannot carry 
the long-range interaction and would be inactive at the infrared scale, 
it seems natural to expect that 
only diagonal gluons are propagated over the large distance 
and carry the long-range confining force in QCD. 
In fact, infrared abelian dominance may be explained 
with the mass generation hypothesis of off-diagonal gluons. 
(This interpretation of infrared abelian dominance 
may be similar to the dominant contribution of photons 
for the large-scale interaction in the Weinberg-Salam model, 
where only the massless photon field is propagated over 
the large distance and the massive weak bosons 
does not contribute to the large-range interaction.)

Since the effective mass $\Meff$ is closely related to 
the off-diagonal gluon propagation, especially on the interaction range, 
we study the gluon propagator in the MA gauge 
using the lattice QCD Monte-Carlo simulation. 
In the following sections, 
we aim to estimate the effective mass of the off-diagonal gluon from 
the numerical result of the gluon propagator in the MA gauge 
\cite{Sgnmykis,AS}. 

\section{Massive Vector Boson Propagator}
In this section, we try to understand the 
behavior of the massive vector boson propagator 
in the Euclidean metric from a simple model, 
for the preparation of the analysis 
on the effective gluon mass in the MA gauge in the lattice QCD.
We start from the Lagrangian of 
the free massive vector field $A_\mu$ with the mass $M \ne 0$ 
in the Proca formalism \cite{itzykson},
\bea
{\cal L}&=& \f{1}{4}(\d_\mu A_\nu - \d_\nu\Amu)^2
	+\f{1}{2}M^2\Amu\Amu,
                                          \label{eqn:Lag} 
\eea
in the Euclidean metric. 
The propagator $\tilde{G}_{\mu\nu}(k;M)$ 
of the massive vector boson is given as 
\bea
\tilde{G}_{\mu\nu}(k;M)
&\equiv& \f{1}{k^2+M^2}
 	\left( \delta_{\mu\nu}+\f{\kmu\knu}{M^2}\) 
\eea
in the momentum representation \cite{MunstrLttcetxt}.
The Euclidean propagator ${G}_{\mu\nu}(x-y;M)$ in the coordinate space is 
obtained by performing the Fourier transformation as
\bea
{G}_{\mu\nu}(x-y;M) \equiv \la \Amu(x) A_\nu(y) \ra
	&=&
	\int \f{d^4 k}{(2\pi)^4} e^{i k \cdot (x-y)} 
		\tilde{G}_{\mu\nu}(k;M).  
\eea

For the investigation of the interaction range, 
it is convenient to examine the scalar-type propagator 
${G}_{\mu\mu}(r;M)$, 
since ${G}_{\mu\mu}(r;M)$ depends only 
on the Euclidean four-dimensional distance $r \equiv \sqrt{(x_\mu- y_\mu)^2}$. 
As shown in Appendix \ref{sec:app},
the scalar-type propagator $\Gmumu(r;M)$ can be expressed 
with the modified Bessel function $K_1(z)$ as 
\bea
{G}_{\mu\mu}(r;M) &=& \la \Amu(x) \Amu(y) \ra
	=\int\f{d^4 k}{(2\pi)^4} e^{i k \cdot (x-y)}
	 \f{1}{k^2+M^2}
 	\left( 4+\f{k^2}{M^2}\) \no\\
	&=&
	3\int\f{d^4k}{(2\pi)^4} e^{i k \cdot (x-y)}	\f{1}{k^2+M^2}    
	 + \f{1}{M^2}{\delta^4(x-y)} 
	=
	\f{3}{4\pi^2}\f{M}{r}K_1(Mr)+\f{1}{M^2}\delta^4(x-y).
					\label{eqn:prp02}
\eea
In the infrared region with large $Mr$, 
Eq.(\ref{eqn:prp02}) reduces to 
\bea
G_{\mu\mu}(r;M)
	&\simeq&
	\f{3\sqrt{M}}{2(2\pi)^{\f{3}{2}}} \f{e^{-Mr}}{r^\f{3}{2}}, 
						\label{eqn:prp03} 
\eea
using the asymptotic expansion 
\bea
K_1(z) \simeq
	\sqrt{\f{\pi}{2z}} e^{-z} 
	\sum^\infty_{n=0}
	\f{\Gamma(\f{3}{2}+n)}{n!\Gamma(\f{3}{2}-n)} \f{1}{(2z)^n} 
						\label{eqn:prp03A}
\eea
for large ${\rm Re} z$. 
Here, the Yukawa-type damping factor $e^{-Mr}$ 
expresses the short-range interaction in the coordinate space.
Then,
the mass $M$ of the vector field $\Amu(x)$ is estimated 
from the slope in the logarithmic plot of 
$\f{r^\f{3}{2}}{\sqrt M }G_{\mu\mu}(r;M)$ as 
the function of $r$, 
\bea
\ln \left\{ {\f{r^\f{3}{2}}{\sqrt M }G_{\mu\mu}(r;M)} \right\} 
	&\simeq& 
	{-Mr}+{\rm const.}
					\label{eqn:prp04}
\eea
This approximation seems applicable for 
$Mr > 1$ from the numerical calculation \cite{KA_HS_prep}. 

For the massless free vector boson, 
the Euclidean scalar-type propagator in the Landau gauge is written as 
\bea
\Gmumu(r) \equiv \f{3}{4\pi^2}\f{1}{r^2} \label{eqn:prp05}
\eea
with the four-dimensional distance $r \equiv \sqrt{(x_\mu- y_\mu)^2}$. 
This exhibits the Coulomb-type interaction 
in the four-dimensional coordinate space. 

\section{Maximally Abelian Gauge Fixing}
\label{section:mag}
In this section, we review the maximally abelian (MA) gauge, 
which seems the key concept for the study of 
the dual-superconductor picture from the lattice QCD 
[18,20-26]. 
In the Euclidean QCD, the MA gauge is defined by minimizing 
the global amount of the off-diagonal gluon, 
\bea
\RchA
	&\equiv&
	\int d^4x \tr 	\left\{ [\hat{D}_\mu,\vec{H}]
			      [\hat{D}_\mu,\vec{H}]^\dagger 
			\right\}
	= \f{e^2}{2} \int d^4x \sum^{N_c^2-N_c}_{\alpha=1}|\Aalmu|^2, 
				\label{eqn:mag001}
\eea
by the SU($N_c$) gauge transformation 
\cite{ichiemp,Sgnmykis,ichiead,HS_HI_etal_adelaid}. 
Here, we have used the Cartan decomposition, 
$\Amu(x)\equiv \vAmu(x)\cdot \vH + \sum^{N_c^2-N_c}_{\alpha=1}\Aalmu (x)E^{\alpha}$ 
and 
the covariant derivative operator $\hat{D}_{\mu} \equiv \hat{\d}_\mu + ie\Amu$.
In the MA gauge, the SU($N_c$) gauge symmetry is reduced into 
the U(1)$^{N_c-1}$ gauge symmetry with the global Weyl symmetry 
%\cite{Sgnmykis,ichiead,HS_MF98,HI_etalPRD96}, 
[18,22,32-34].
and the diagonal gluon component $\vAmu$ 
behaves as the abelian gauge field 
and the off-diagonal gluon component $\Amu^\alpha$
behaves as the charged matter field in terms of the 
residual U(1)$^{N_c-1}$ abelian gauge symmetry.
In the MA gauge, off-diagonal gluon components are forced 
to be as small as possible, and then the gluon field $\Amu(x)$ 
mostly approaches to the abelian gauge field 
$\vAmu(x)\cdot \vH $ \cite{ichiead}.

Since the covariant derivative operator 
$\hat{D}_\mu \equiv \hat{\d}_\mu + ie\Amu$ obeys 
the adjoint transformation 
$\hat D_\mu \rightarrow \Omega  \hat D_\mu  \Omega ^\dagger$,
it is convenient to describe $\RchA$ 
with $\hat{D}_\mu$ like Eq.(\ref{eqn:mag001}), 
for the argument of the gauge transformation property of $\RchA$.
Using the infinitesimal gauge transformation, 
the local gauge-fixing condition in the MA gauge is obtained as
\bea
[ \vH ,[\hDdmu,[\hDdmu,\vH ] ] ]=0.			  \label{eqn:201}
\eea
In particular for the SU(2) QCD, 
this leads to $\( i\d_\mu \pm e\Amu^{3} \) A^{\pm}_{\mu} = 0$,  
or equivalently, the local variable 
$
\Phi_{\rm MA}[\Amu(x)]\equiv[\hDdmu,[\hDdmu,\tau^3 ]]
 					\label{eqn:Rfinala}
$
is diagonalized in the MA gauge.

In the SU(2) lattice QCD, the system is described by the link variable 
$\Umu(s) = \exp \{ iae\Amua(s)\f{\tau^a}{2} \} \in$ SU(2) 
with the lattice spacing $a$ and the QCD gauge coupling constant $e$.
The MA gauge is defined by maximizing
\bea
R_{\rm MA}
	&\equiv&
	 \sum_s \sum^4_{\mu =1} 
	\tr \biggr\{ \Umu(s) \tau^3 \Umu^{\dagger}(s) \tau^3  \biggr\}
						\label{eqn:mag002}
\eea
using the SU(2) gauge transformation.
In the continuum limit $a \rightarrow 0$,
maximizing $R_{\rm MA}$ 
leads to minimizing $\RchA$ in Eq.(\ref{eqn:mag001}).
Also in the lattice formalism, 
the MA gauge fixing is a partial gauge fixing on the coset space
SU(2)$^{\rm local}$/(U(1)$_3^{\rm local} \times $ Weyl$^{\rm global}$),
and there remain the U(1)$_3$-gauge symmetry and the global Weyl symmetry.
According to the partial gauge fixing on SU(2)/U(1)$_3$, 
it is convenient to use the Cartan decomposition 
for the SU(2) link-variable as $\Umu(s)\equiv\Mmu(s)\umu(s)$
with $\umu(s) \equiv e^{i\theta_\mu^3(s)\tau^3} \in $ U(1)$_3$ and
$\Mmu(s) \equiv e^{i \left\{ \theta_\mu^1(s)\tau^1 + \theta_\mu^2(s)\tau^2 \right\} }
\in$ SU(2)/U(1)$_3$.
Here,  the abelian link variable $\umu(s)$,
which is suggested to play the relevant role for confinement 
[18,20-22] 
in the MA gauge, 
obeys the residual U(1)$_3$-gauge transformation,
\bea
\umu(s)\rightarrow \omega(s)\umu(s)\omega^{\dagger}(s+\hat{\mu}),
						\label{eqn:GT}
\eea
with the gauge function $\omega(s) \in$ U(1)$_3$.

For the study of the gluon propagator $\la\Amu^a(x) A_\nu^b(y)\ra$
$(a,b=1,2,3)$ in the MA gauge, 
the gluon field $\Amua(x)$ itself is to be defined 
as the dynamical variable by removing
the residual U(1)$_3$-gauge degrees of freedom.
In order to extract the most continuous gluon configuration
under the constraint of the MA gauge fixing, 
we take the U(1)$_3$ Landau gauge 
for the U(1)$_3$-gauge degrees of freedom 
remaining in the MA gauge.
The U(1)$_3$ Landau gauge is defined by maximizing
\bea
R_{\rm U(1)L}\equiv \sum_s \sum_{\mu=1}^4 \tr[u_\mu(s)]
											\label{eqn:defU1Landau}
\eea
using the U(1)$_3$-gauge transformation (\ref{eqn:GT}) 
[18,27,34-37]. 
The maximizing condition for $R_{\rm U(1)L}$
leads to the Landau gauge condition
$\d_\mu \Amu^3=0$ in the continuum limit.
By taking the U(1)$_3$ Landau gauge, 
all of the abelian link variable $\umu(s)$ mostly approach to unity,
and the most continuous gluon configuration 
is realized under the MA gauge fixing condition.

\section{Gluon Propagators and Off-diagonal Gluon Mass in the MA gauge in the Lattice QCD}
\subsection{Gluon Propagators on Lattices}
In this sub-section, we consider the extraction of the gluon propagator
from the link variable $\Umu(s)$
in the MA gauge with the U(1)$_3$-Landau gauge fixing.
To begin with, we extract the lattice gluon field $\Amua(s)~ (a=1,2,3)$ 
from the link variable $\Umu(s)$
obtained in the lattice QCD Monte Carlo simulation.
In the SU(2) lattice QCD, $\Umu(s)$ relates to $\Amu(s)$ as
\bea
\Umu(s)	&\equiv& 
{\cal P}\exp \( ie \int_s^{s+a\hat{\mu}} dx_\mu \Amu^a(x) \f{\tau^a}{2} \)
=\exp\left\{i ae \overline{\Amua}(s)\f{\tau^a}{2}\right\}~(a=1,2,3)
					\label{eqn:AAform01}
\eea
with a small lattice spacing $a$, the gauge coupling constant $e$ 
and the pass-ordering operator $\cal P$.
Here, $\overline{\Amua}(s)$ corresponds to 
the coarse-grained field of $\Amua(x)$  
on the link from $s$ to $s+a\hat{\mu}$, 
and coincides with $\Amu^a(s)$ approximately, 
when $a$ is enough small\cite{Rothe}.
In this paper, 
we regard $\overline{\Amua}(s)$ as the approximate gluon field 
on the lattice.
Here, the gluon field $\overline{\Amua}(s)$ is described as
\bea
\overline{\Amua}	&=&
	\f{2}{ae}\f{\Umu^a}{\sqrt{\sum^3_{a=1}(\Umu^a)^2}}
	\arctan\f{\sqrt{\sum^3_{a=1}(\Umu^a)^2}}{\Umu^0}
						\label{eqn:AAform02}
\eea
by parameterizing $\Umu(s)$ as
$
\Umu	\equiv 
	\Umu^0 + i \tau^a \Umu^a~~
		(\Umu^0,\Umu^a \in {\bf R}; ~a=1,2,3) 
$
, which satisfies the unitarity condition,
$
(\Umu^{0})^{2}+(\Umu^{a})^{2}=1.
$
The gluon field $\overline{\Amu}(s) \in {\rm su(2)}$ is thus obtained from
the link variable $\Umu(s) \in {\rm SU(2)}$ generated in the lattice QCD 
Monte Carlo simulation.
For the simple notation, we express $\overline{\Amua}(s)$ by  
$\Amua(s)$ hereafter.

Next, 
we study the Euclidean gluon propagator $G_{\mu \nu}^{ab}(x-y)=
\langle A_\mu^a(x) A_\nu^b(y) \rangle$ 
in the MA gauge with the U(1)$_3$-Landau gauge.
In this gauge, $\Umu(s)$ is determined without the ambiguity 
on the local gauge transformation.
In this paper, we study the Euclidean scalar-type propagator of 
the diagonal (neutral) gluon as
\bea
\Gmumu^{\rm Abel}(r) \equiv \la \Amu^3(x)\Amu^3(y)\ra,
				\label{eqn:AAf002}
\eea
and that of the off-diagonal (charged) gluon as
\bea
\Gmumu^{\rm off}(r) \equiv \la \Amu^+(x)\Amu^-(y)\ra 
	 &=&
\f{1}{2} \left\{ \la \Amu^1(x)\Amu^1(y)\ra + \la \Amu^2(x)\Amu^2(y)\ra 
\right\}
				\label{eqn:AAf003}
\eea
with the off-diagonal gluons 
$\Amu^{\pm}(x)\equiv\f{1}{\sqrt{2}} \left\{ \Amu^1(x) \pm i\Amu^2(x) \right\}$.
These scalar-type propagators are expressed  
as the function of the four-dimensional Euclidean distance
$r\equiv\sqrt{ (x_\mu- y_\mu)^2 }$.
In $\Gmumu^{\rm off}(r)$, the imaginary part  
$-\f{i}{2}\left\{ \la \Amu^1(x)\Amu^2(y) \ra 
-\la \Amu^2(x)\Amu^1(y) \ra \right\}$  
disappears automatically due to the symmetry as 
$\la \Amu^1(x)\Amu^2(y)\ra = \la \Amu^2(x)\Amu^1(y)\ra $.

\subsection{Lattice QCD Results for Gluon Propagators}
Using the SU(2) lattice QCD,
we calculate the gluon propagator in the MA gauge 
with the U(1)$_3$ Landau gauge fixing.
The  Monte Carlo simulation is performed 
on the $12^3 \times 24$ lattice with 
$2.2 \le \beta \le 2.4$, 
and also on the $16^4$ and $20^4$ lattices with 
$2.3 \le \beta \le 2.4$. 
All measurements are done every 500 sweeps after a thermalization of 
10,000 sweeps using the heat-bath algorism.
We prepare 100 gauge configurations at each $\beta$ for the calculation.

The MA gauge fixing and the U(1)$_3$ Landau gauge fixing are performed 
by the iteration of the local maximization of $R_{\rm MA}$ and 
$R_{\rm U(1)L}$, respectively, 
using the overrelaxation algorithm with 
the overrelaxation parameter $\omega=1.7$\cite{MO90}.
As the accuracy of the maximizing condition of 
$R_{\rm MA}$ and $R_{\rm U(1)L}$,  
they are required to converge to $10^{-7}$ per a sweep. 

The physical scale unit is determined so as to reproduce 
the string tension as $\sigma=0.89$ GeV/fm $\simeq(420$ MeV$)^2$ 
based on Refs. \cite{CMicMTep,UKQCD93,CMicSPer,JFinUHelFKar,BalCSchKSch}.
In Fig.{\ref{caption:sqsigma}} and Table 1
\cite{CMicMTep,UKQCD93,CMicSPer,JFinUHelFKar,BalCSchKSch}, 
we summarize the SU(2) lattice QCD data for $\sqrt{\sigma} a$, 
the square root of the string tension in the lattice unit,  
as the function of $\beta$ in the case of the   
pure gauge standard action. 
In the region of $2.2 \le \beta \le  2.4 $, 
the lattice QCD data in Fig.{\ref{caption:sqsigma}} 
seem to lie on a straight line, and therefore  
we determine the values of $\sqrt{\sigma} a$ and the scale unit $a$ 
at intermediate values of $\beta \in (2.2, 2.4)$
using the interior division.

The Euclidean scalar-type propagators $\Gmumu^{\rm Abel}(r)$ and 
$\Gmumu^{\rm off}(r)$ 
are measured as the two point functions of the gluon fields on 
$(\vec{x}, 0)$ and $(\vec{x}, r)$ 
using Eqs.(\ref{eqn:AAf002}) and (\ref{eqn:AAf003}).
While the temporal direction is fixed along the longest side
in the $12^3 \times 24$ lattice, 
we measure the correlation in all four direction 
for the symmetric lattices ($16^4$ and $20^4$) 
without fixing the temporal direction. 
We show in Fig.\ref{caption:2} the lattice QCD data for 
the diagonal gluon propagator $\Gmumu^{\rm Abel}(r)$ 
and the off-diagonal gluon propagator $\Gmumu^{\rm off}(r)$ 
in the MA gauge with the U(1)$_3$ Landau gauge fixing.  
Here, error is estimated with the jackknife 
analysis \cite{MunstrLttcetxt}. 
In the MA gauge, $\Gmumu^{\rm Abel}(r)$ and 
$\Gmumu^{\rm off}(r)$ manifestly differ .
The diagonal-gluon propagator $\Gmumu^{\rm Abel}(r)$ 
takes a large value even at the long distance. 
In fact, the diagonal gluon $\Amu^3$ in the MA gauge 
propagates over the long distance.
On the other hand, the off-diagonal gluon propagator 
$\Gmumu^{\rm off}(r)$ rapidly decreases and is negligible
for $r \gsim 0.4$ fm in comparison with $\Gmumu^{\rm Abel}(r)$. 
Then, the off-diagonal gluon $\Amu^\pm$ seems to propagate 
only within the short range as $r \lsim 0.4$ fm.
Thus, `infrared abelian dominance' for the gluon propagator 
is found in the MA gauge.

Finally in this subsection, we briefly discuss the finite size effect 
on the gluon propagators, $\Gmumu^{\rm Abel}$ and $\Gmumu^{\rm off}$.
As shown in Fig.{\ref{caption:2}}, the shape of 
$\Gmumu^{\rm Abel}$ slightly depends on the lattice parameters 
($\beta$ and the lattice size), although $\beta$-dependence of 
$\Gmumu^{\rm Abel}$ at $\beta \in [2.3, 2.4]$ seems to decrease  
for the $20^4$ lattice. (See also Figs.3 and 4.) 
Then, one should be careful about the finite-size effect 
for the detailed argument on the diagonal gluon propagator 
$\Gmumu^{\rm Abel}$.
On the other hand, the single curve approximately fits 
almost all the lattice data of $\Gmumu^{\rm off}$ obtained  
 from different lattices on $\beta$ and the size. 
Therefore, the finite-size effect seems to be small on 
the off-diagonal gluon propagator $\Gmumu^{\rm off}$ in the MA gauge.
(See also Figs.3-5.)

\subsection{Estimation of the Off-diagonal Gluon Mass in the MA Gauge}

In this subsection, we examine the hypothesis on effective mass 
generation of off-diagonal gluons in the MA gauge.
To this end, 
we investigate the logarithm plot of $r^{3/2}\Gmumu^{\rm off}(r)$ 
as the function of $r$,
since the massive vector boson with the mass $M$ behaves like
$\Gmumu(r)\sim\f{\exp (-Mr)}{r^{3/2}}$ 
for $Mr > 1$ \cite{KA_HS_prep}  
as was discussed in Section 3.

In Fig.3, 
we show the logarithmic plot of  
$r^{3/2}\Gmumu^{\rm off}(r)$ and $r^{3/2}\Gmumu^{\rm Abel}(r)$
as the function of the distance $r$
in the MA gauge with the U(1)$_3$ Landau gauge fixing.  
As shown in Fig.3, the logarithmic plot of 
$r^{3/2}\Gmumu^{\rm off}(r)$
seems to decrease linearly in the region of $0.2 {\rm~fm} \lsim r \lsim 1$ fm.
In fact, $\Gmumu^{\rm off}(r)$ behaves as the Yukawa-type function 
\bea
\Gmumu^{\rm off}(r)\sim\f{\exp(-\Meff~r)}{r^{3/2}}
\eea
for $0.2 {\rm~fm} \lsim r \lsim 1$ fm in the MA gauge.
 From the linear slope on $r^{3/2}\Gmumu^{\rm off}(r)$ in Fig.3, 
the effective off-diagonal gluon mass $\Meff$ is roughly estimated as
\bea
\Meff \simeq 5 ~\mbox{fm}^{-1} \simeq 1~\mbox{GeV}. 
					\label{eqn:chmass}
\eea

For more accurate estimation of the off-diagonal gluon mass $\Meff$, 
we analyze the lattice data of $r^{3/2}\Gmumu^{\rm off}(r)$ 
individually at each lattice with the same $\beta$ and the lattice size, 
to separate the systematic error relating to the lattice parameters.
We show the lattice data for $r^{3/2}\Gmumu^{\rm off}(r)$ 
in Figs.4 (a),(b) and (c) corresponding to the lattice size, 
$12^3 \times 24$, $16^4$ and $20^4$, respectively. 
Using the least squares method, 
we perform the linear fitting analysis for 
the logarithmic plot of $r^{3/2}\Gmumu^{\rm off}(r)$ 
with the fitting range of $0.2 {\rm~fm} \le r \le 1 {\rm~fm}$.

We summarize in Table 2 
the effective off-diagonal gluon mass $\Meff$ 
obtained from the slope analysis 
at each lattice (the lattice size and  $\beta$) 
as well as $\chi^2$ and the degrees of freedom, $N_{df}$. 
The error of $\Meff$ is estimated with the jackknife analysis.
Here, the two crude-lattice data with $12^3 \times 24$ and 
$\beta=$ 2.2, 2.23 show extremely large values on $\chi^2 /N_{df}$, 
which means the wrong fitting of these two data, 
and hence the errors listed are not creditable for 
the two crude-lattice data. 
Except for these two wrong-fitting data, 
the lattice QCD data for the off-diagonal gluon mass in the MA gauge 
distribute around $\Meff \simeq 1.2{\rm~GeV}$ 
within about 0.1 GeV error. 

We show in Fig. 5 the off-diagonal gluon mass $\Meff$ as the function 
of $1/V$, the inverse volume in the physical unit, 
to investigate the finite-size effect on $\Meff$. 
In Fig.5, we have dropped the two wrong-fitting data 
on the crude lattice with $12^3 \times 24$ and $\beta=$ 2.2, 2.23. 
The systematic errors including the finite-size effect seem to be 
small for $\Meff$. 

Finally in this section, we discuss the relation 
between infrared abelian dominance and 
the off-diagonal gluon mass $\Meff \simeq 1.2$ GeV, 
which is considered to be induced by the MA gauge fixing. 
Due to the large effective mass $\Meff$, 
the off-diagonal gluon propagation is restricted within about
$\Meff^{-1} \simeq 0.2$ fm in the MA gauge.
Therefore, at the infrared scale as $r \gg 0.2$ fm,
the off-diagonal gluons $\Amu^\pm$ cannot mediate the long-range force, 
and only the diagonal gluon $\Amu^3$ can mediate  
the long-range interaction in the MA gauge.
In fact, in the MA gauge, the off-diagonal gluon is expected to be 
inactive due to its large mass $\Meff$ in the infrared region 
in comparison with the diagonal gluon. 
Then, we conjecture that effective mass generation of 
the off-diagonal gluon in the MA gauge would 
be an essence of abelian dominance in the infrared region.

\section{Summary and Concluding Remarks}
In this paper, we have studied the gluon propagator 
and the off-diagonal gluon mass in the MA gauge  
with the U(1)$_3$ Landau gauge fixing using the SU(2) lattice QCD.
The Monte Carlo simulation is performed 
on the $12^3 \times 24$ lattice with $2.2 \le \beta \le 2.4$, 
and also on the $16^4$ and $20^4$ lattices with 
$2.3 \le \beta \le 2.4$. 
In the MA gauge, we have measured the Euclidean scalar-type 
propagators $G_{\mu\mu}(r)$ of the diagonal and the off-diagonal gluons, 
and have found infrared abelian dominance for the gluon propagator. 
In the MA gauge, we have found that 
the off-diagonal gluon behaves as a massive vector boson 
with the effective mass $M_{\rm off} \simeq 1$ GeV for $r \gsim 0.2$ fm. 
At each lattice, we have estimated the effective off-diagonal gluon mass 
$M_{\rm off}$ from the linear fitting analysis of the logarithmic 
plot of $r^{3/2} G^{\rm off}_{\mu\mu}(r)$. 
The off-diagonal gluon mass is estimated as $\Meff \simeq 1.2$ GeV 
with the error about 0.1 GeV. 
We have also checked the smallness of the finite-size effect for 
the off-diagonal gluon mass $\Meff$ in the MA gauge, 
but still now, we have to consider the finite lattice-spacing effect 
and the Gribov copy effects \cite{Huang,BaSchlSchi,BorMitMulPahl} 
relating to the gauge fixing for more quantitative arguments.

From the behavior of the diagonal-gluon propagator $\Gmumu^{\rm Abel}(r)$,
the diagonal gluon seems to behave as a light or massless particle.
However, for the detailed argument on $\Gmumu^{\rm Abel}(r)$,
one should consider the finite size effect more carefully,
because the diagonal gluon may propagate over the long distance
beyond the lattice size.
On the other hand, the off-diagonal gluon propagation is limited within
the short distance as $r \lsim \Meff^{-1} \simeq 0.2$ fm,
and hence the finite size effect for $\Gmumu^{\rm off}$ is 
expected to be small enough,
when the lattice size is taken to be much larger than
$\Meff^{-1} \simeq 0.2$ fm.
Such a tendency of the small finite-size effect on $\Gmumu^{\rm off}$ 
has been checked using the lattice QCD simulation 
with the various volume. 

Due to the large off-diagonal gluon mass as $\Meff \simeq 1.2$ GeV, 
the off-diagonal gluon cannot mediate 
the interaction over the large distance as $r \gg M_{\rm off}^{-1}$, 
and such an infrared inactivity of the off-diagonal gluon may 
lead infrared abelian dominance in the MA gauge.
Then, an essence of infrared abelian dominance would be 
physically explained as generation of the off-diagonal gluon mass 
$M_{\rm off}$ induced by the MA gauge fixing.
As an interesting conjecture, 
this off-diagonal gluon mass generation may provide  
general abelian dominance 
for the long-distance physics in QCD in the MA gauge, 
similar to the infrared inactivity of the massive 
weak-bosons in the Weinberg-Salam model.

Finally, we consider the criterion of the scale where 
abelian dominance would hold in the MA gauge. 
We conjecture that $\Meff^{-1} \simeq 0.2$ fm would be regarded as 
the critical scale on abelian dominance in the MA gauge.
For the short distance as $ r \lsim M_{\rm off}^{-1} \simeq 0.2$ fm, 
the effect of off-diagonal gluons becomes significant, 
and hence all the gluon components have to 
be considered even in the MA gauge. 
On the other hand, 
at the long distance as $r \gg M_{\rm off}^{-1} \simeq 0.2$ fm, 
the off-diagonal gluon cannot mediate the interaction and would be 
negligible in comparison with the diagonal gluon, 
which may lead abelian dominance in the MA gauge. 
In this way, abelian dominance is expected to hold 
for the long-range physics as $r \gg \Meff^{-1}$, 
and abelian dominance would break at the short distance as 
$r \lsim \Meff^{-1}$ in QCD in the MA gauge.

~\\~\\
{\bf Acknowledgements}\\
We would like to thank Drs. H. Matsufuru and A. Tanaka 
for their useful comments and discussions. 
We would like to thank all members of the RCNP theory group.
One of authors (H.S.) is supported in part by Grant for 
Scientific Research (No.09640359) from the Ministry of Education,
Science and Culture, Japan.
The lattice QCD simulations have been performed on SX-4 at RCNP, Osaka.

\appendix
\section{Formula on Euclidean  Propagator}
\label{sec:app}
In this appendix, relating to Eqs.(\ref{eqn:prp02}) and (\ref{eqn:prp05}), 
we add several formula on the vector boson propagator in the 
Euclidean metric.
For the massive vector case, 
the scalar-type propagator $\Gmumu(r;M) \equiv \la \Amu(x) \Amu(y)  \ra$ 
is expressed as 
\bea
{G}_{\mu\mu}(r;M) 
	&=&
	3\int\f{d^4k}{(2\pi)^4} e^{i k \cdot (x-y)}\f{1}{k^2+M^2}    
	 + \f{1}{M^2}{\delta^4(x-y)}. ~\label{eqn:prp01x}
\eea
Since ${G}_{\mu\mu}(r;M)$ depends only 
on the four-dimensional distance $r \equiv \sqrt{(x_\mu-y_\mu)^2}$,
one takes $x_\mu-y_\mu = (r,0,0,0) $ without loss of generality 
for the calculation of $\Gmumu(r;M)$. 
Then, the integration in Eq.(\ref{eqn:prp01x}) can be
expressed with $K_1(z)$ as 
\bea
	\int\f{d^4k}{(2\pi)^4} e^{i k \cdot (x-y)}
	\f{1}{k^2+M^2}  
	&=&
	\int \f{d^3k}{(2\pi)^3}
	\( \int_{-\infty}^{\infty} 
	\f{dk_0}{2\pi}  e^{i k_0 r} \f{1}{k_0^2+\mbox{\boldmath $k$}^2 +M^2} \) \no\\
	&=&
	\int \f{d^3k}{(2\pi)^3}
	\f{1}{2\sqrt{\mbox{\boldmath $k$}^2+M^2}} 
        e^{-\sqrt{\mbox{\footnotesize \boldmath $k$}^2+M^2} \ r} 
	=
	\f{1}{4\pi^2}\int_0^\infty dk
	\f{k^2}{\sqrt{k^2+M^2}} e^{-\sqrt{k^2+M^2} \ r} \no \\
	&=&
	\f{1}{4\pi^2}\int_M^\infty dE e^{-Er}
	\sqrt{E^2-M^2} 
	=
	\f{1}{4\pi^2} M^2 \int_1^\infty d \epsilon e^{- \epsilon Mr}
	\sqrt{\epsilon^2-1} \no \\
	&=&
	\f{1}{4\pi^2}\f{M}{r} K_1(Mr)
\eea
with $E \equiv \sqrt{\mbox{\boldmath $k$}^2 + M^2}$ and $\epsilon \equiv E/M$.
Here, we have used the integration formula for the modified Bessel function,
\bea
K_1(z) &\equiv& 
	z\int_1^\infty dt e^{-zt} \sqrt{t^2-1} ~~~~~~(\mbox{\Fr R} z > 0). 
\eea
Thus, the scalar-type propagator $\Gmumu(r;M)$ can be expressed as
\bea
G_{\mu\mu}(r;M)
	&=&
	\f{3}{4\pi^2}\f{M}{r}K_1(Mr)+\f{1}{M^2}\delta^4(x-y).
					\label{eqn:prp02ap}
\eea

For the massless vector case,
we use the covariant Lagrangian in the Euclidean metric,
\bea
{\cal L}&=& \f{1}{4} \( \d_\mu A_\nu - \d_\nu\Amu \) ^2
		+\f{1}{2\xi} (\d_\mu A_{\mu})^2   \label{eqn:LagM0} 
\eea
with the gauge-fixing parameter $\xi$.
The massless vector-boson propagator is obtained as
\bea
\tilGmunu(k)&\equiv& \f{1}{k^2} 
 	\biggr\{ \delta_{\mu\nu}+ (\xi-1)\f{\kmu\knu}{k^2}\biggr\}~\label{eqn:M0moG}
\eea
in the momentum representation \cite{MunstrLttcetxt}.
The scalar-type propagator $\Gmumu(x)$ in the coordinate space 
is written as
\bea
\Gmumu(r) \equiv \f{3+\xi}{4\pi^2}\f{1}{r^2},\label{eqn:prp05x}
\eea
which becomes Eq.(\ref{eqn:prp05}) for the Landau gauge $\xi=0$.

\newpage
\begin{figure}
\caption{
The square root of the string tension in the lattice unit,  
$\sqrt{\sigma} a$, 
as the function of $\beta$ in the SU(2) lattice QCD 
with the pure-gauge standard action, using the data in Refs.[38-42].
In the region of $2.2  \le \beta \le 2.4$, the lattice QCD data 
seem to lie on a straight line. 
}
\label{caption:sqsigma}
\end{figure}

\begin{figure}
\caption
{
The SU(2) lattice QCD results for the scalar-type gluon propagators 
$\Gmumu^{\rm Abel}(r) \equiv \la \Amu^3(x)\Amu^3(y)\ra$ and 
$\Gmumu^{\rm off}(r) \equiv \la \Amu^+(x)\Amu^-(y)\ra$
as the function of $r\equiv \sqrt{(x_\mu-y_\mu)^2}$
in the MA gauge with the U(1)$_3$ Landau gauge fixing in the physical unit.
The Monte Carlo simulation is performed 
on the $12^3 \times 24$ lattice with $2.2 \le \beta \le 2.4$, 
and also on the $16^4$ and $20^4$ lattices with 
$2.3 \le \beta \le 2.4$. 
The diagonal-gluon propagator $\Gmumu^{\rm Abel}(r)$ takes 
a large value even at the long distance.
On the other hand, the off-diagonal gluon propagator $\Gmumu^{\rm off}(r)$
rapidly decreases and is negligible for $r \gsim 0.4$ fm 
in comparison with $\Gmumu^{\rm Abel}(r)$.
						\label{caption:2}
} 
\end{figure}

\begin{figure}
%$\hspace{2.6cm}$\epsfig{figure=r3i2AAdir-r.EPSF,height=8cm}
\caption
{ 
The logarithmic plot of 
$r^{\f 3 2}\Gmumu^{\rm off}(r)$ and 
$r^{\f{3}{2}}\Gmumu^{\rm Abel}(r)$
as the function of the distance $r$ 
in the MA gauge with the U(1)$_3$ Landau gauge fixing,
using the SU(2) lattice QCD with 
$12^3 \times 24$ ($2.2 \le \beta \le 2.4$),
$16^4$ and $20^4$ ($2.3 \le \beta \le 2.4$). 
The off-diagonal gluon propagator behaves as the Yukawa-type function 
$\Gmumu^{\rm off}(r)\sim \f{\exp(-\Meff r)}{r^{3/2}}$ 
with $\Meff \simeq 1$ GeV for $r \gsim 0.2$ fm. Therefore, the off-diagonal 
gluon seems to have a large mass $\Meff \simeq 1$ GeV in the MA gauge.
                                                    \label{caption:r2graphABC}
}
\end{figure}

\begin{figure}
\caption
{
The logarithmic plot of 
${r^{\f 3 2}}\Gmumu^{\rm off}(r)$ and 
$r^{\f{3}{2}} \Gmumu^{\rm Abel}(r)$
in the SU(2) lattice QCD with   
(a) $12^3 \times 24$ and $2.2 \le \beta \le 2.4$,
(b) $16^4$ and $2.3 \le \beta \le 2.4$, 
(c) $20^4$ and $2.3 \le \beta \le 2.4$  
in the MA gauge with the U(1)$_3$ Landau gauge fixing. 
The slope corresponds to the effective mass, and 
the effective off-diagonal gluon mass 
is estimated as $\Meff \simeq 1.2$ GeV for $r \gsim 0.2$ fm 
both on these lattices.
The finite lattice-size effect seems to be small 
on the off-diagonal gluon propagator.
}
\end{figure}

\begin{figure}
\caption{
The off-diagonal gluon mass $\Meff$ as the function of $1/V$, 
the inverse lattice volume in the physical unit.
The off-diagonal gluon mass $\Meff$ is obtained from the slope analysis on the 
logarithmic plot of $r^{\f 3 2} \Gmumu^{\rm off}(r)$ at each lattices 
($\beta$ and the lattice size) with the fitting range of $0.2{\rm ~fm}  \le r \le 1{\rm ~fm} $.
In this figure, we drop the two wrong-fitting data on $12^3 \times 24$ with 
$\beta=2.2$ and 2.23, 
because these data show extremely large values on $\chi^2$ (See Table 2).
}
\end{figure}

\newpage
\begin{center}
TABLES
\end{center}

Table 1. 
The SU(2) lattice QCD data for $\sqrt{\sigma} a$,
the square root of the string tension in the lattice unit, 
at various $\beta$ for the pure-gauge standard action. 
These data are taken from Refs.[38-42].
$N_s$ and $N_\tau$ denote the lattice size $N_s^3 \times N_\tau$.
We add the physical length of $a$ to reproduce $\sqrt{\sigma}$=0.89GeV/fm.\\

Table 2. 
The effective off-diagonal gluon mass $\Meff$ 
obtained from the slope analysis on the logarithmic plot of 
${r^{\f 3 2}}\Gmumu^{\rm off}(r)$ at each lattice 
($\beta$ and the lattice size) with the fitting range of 
0.2 fm $\le r \le$ 1.0 fm. Here, $\chi^2$ and the degrees of 
freedom, $N_{df}$, are also listed.
Except for the first two wrong-fitting data with large $\chi^2$, 
one finds 1.1 GeV $\lsim \Meff \lsim$ 1.3 GeV.

~\\
\newpage
\noindent
\begin{figure}
\begin{center}
\epsfig{figure=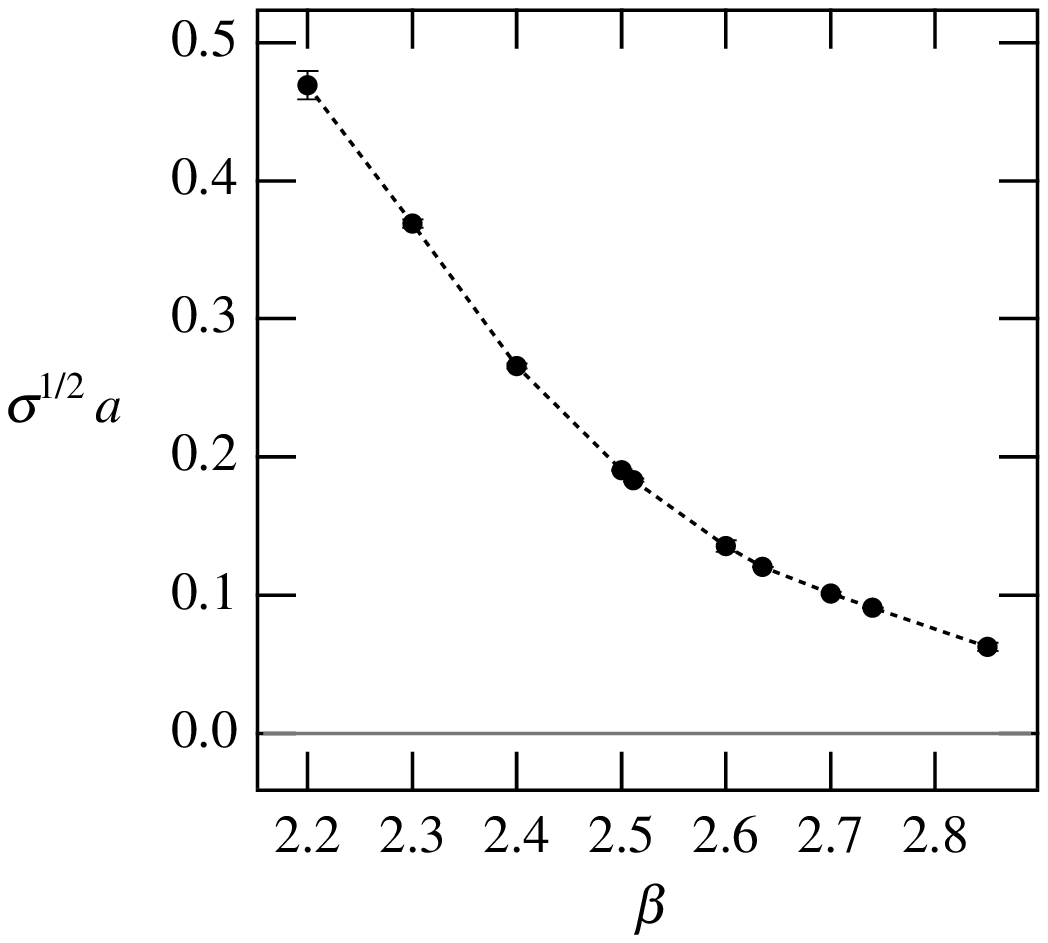}
\\
{\Huge Fig. \ref{caption:sqsigma}}
\end{center}
\end{figure}
\begin{figure}
\begin{center}
\epsfig{figure=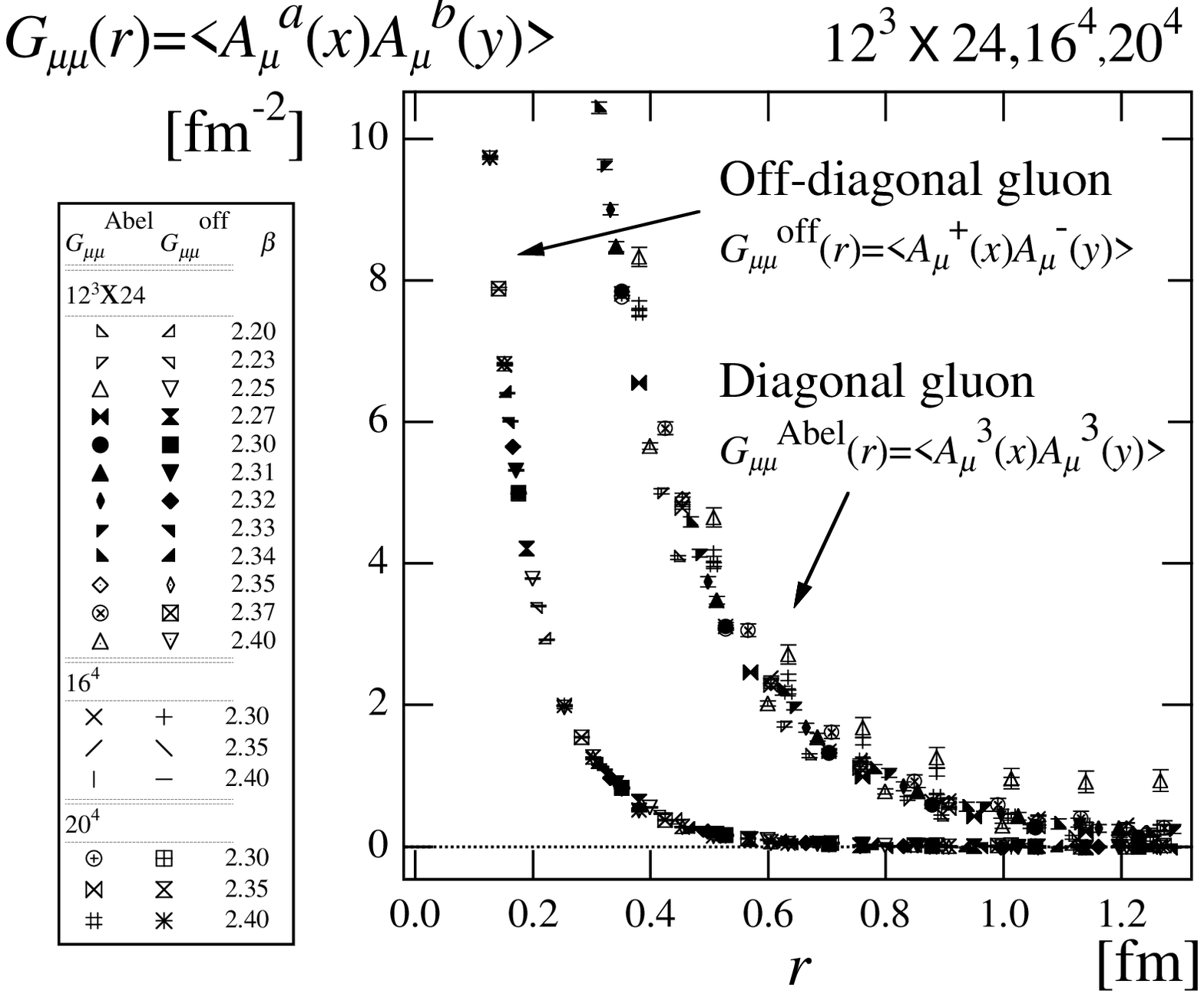} 
\\
{\Huge Fig. \ref{caption:2}}
\end{center}
\end{figure}
\newpage
\begin{figure}
\begin{center}
\epsfig{figure=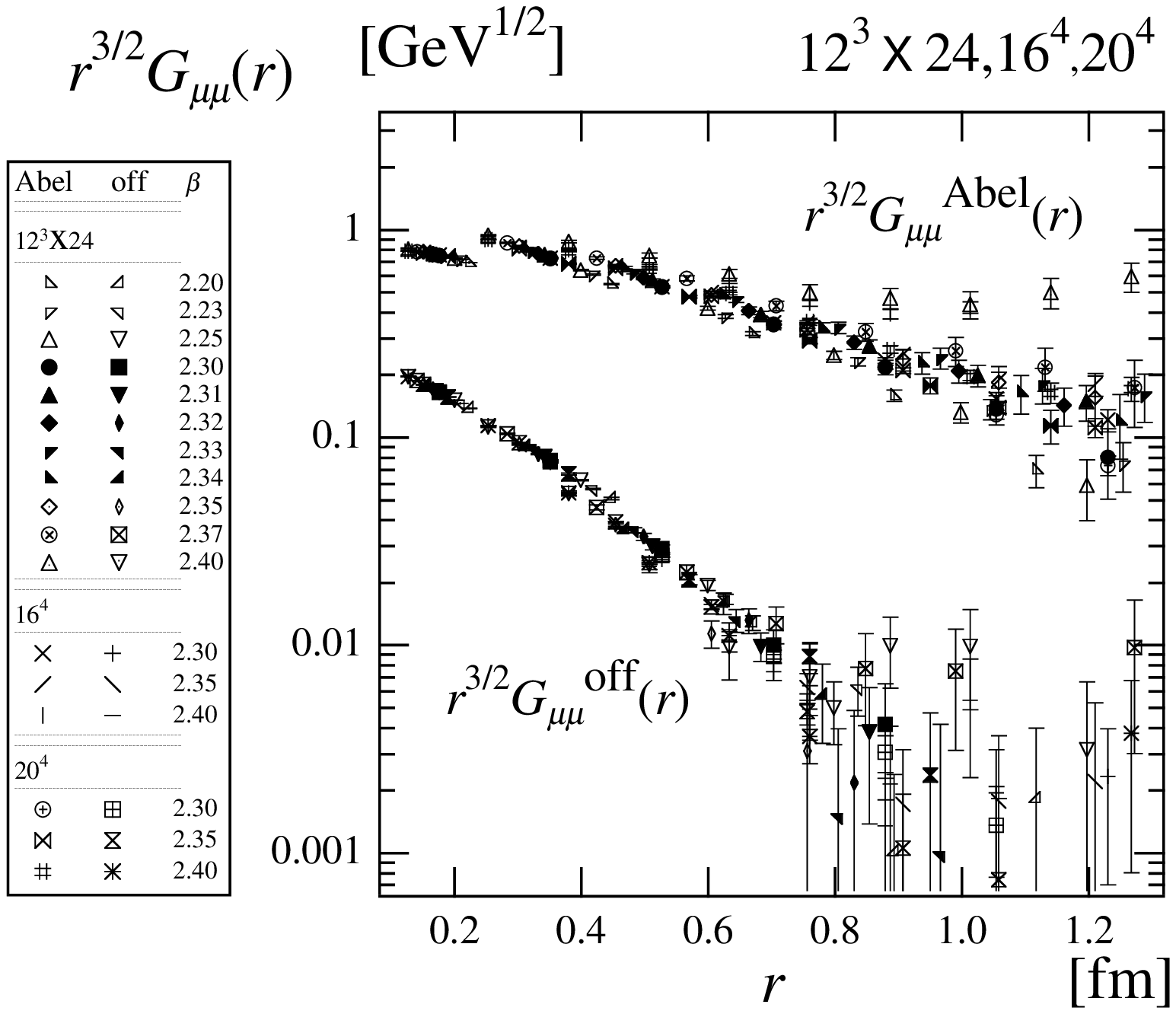} 
~\\
{\Huge Fig. \ref{caption:r2graphABC}}
\end{center}
\end{figure}
\begin{figure}
\begin{center}
\epsfig{figure=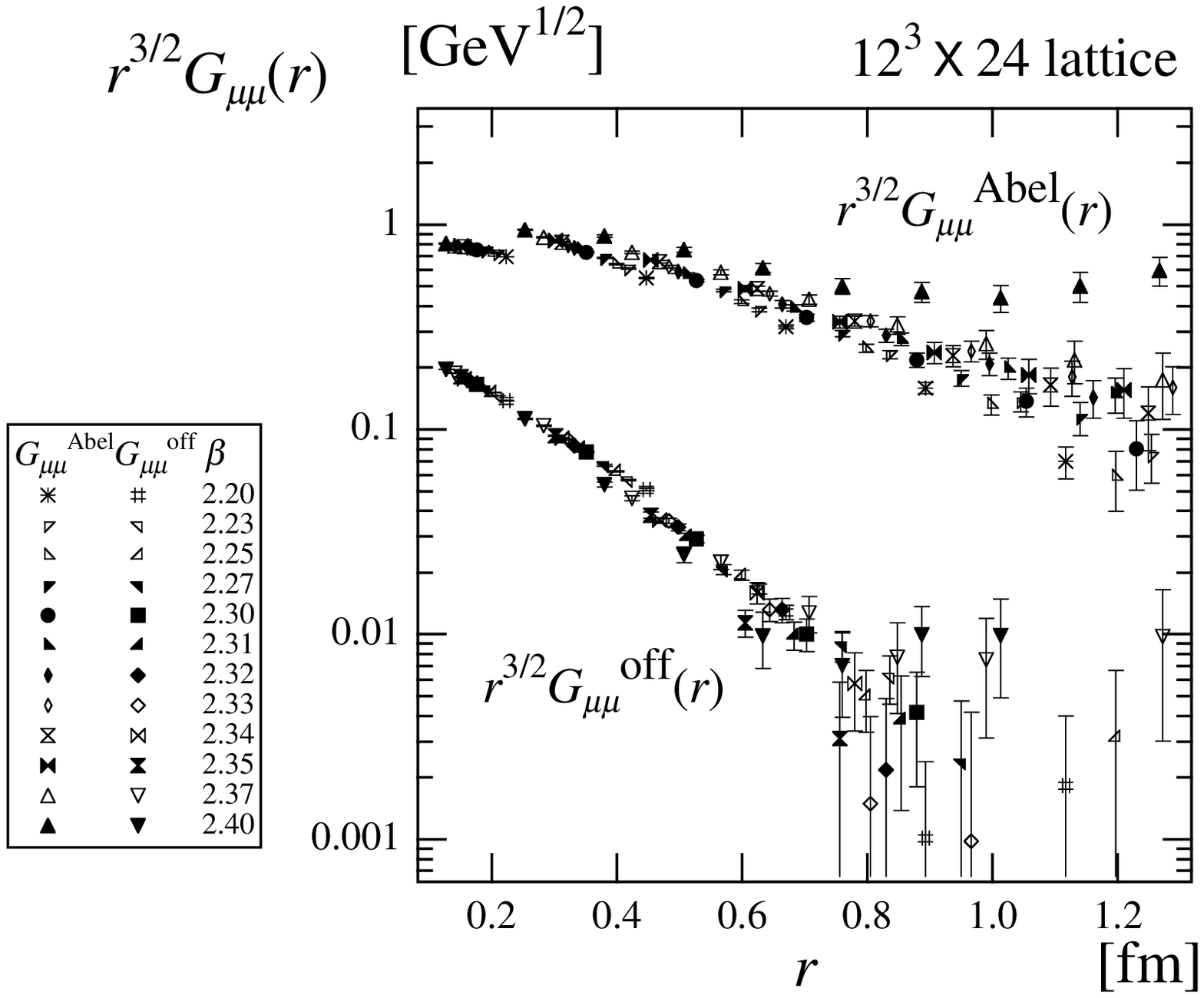} 
~\\
{\Huge Fig. 4(a)} 
\end{center}
\end{figure}
\begin{figure}
\begin{center}
\epsfig{figure=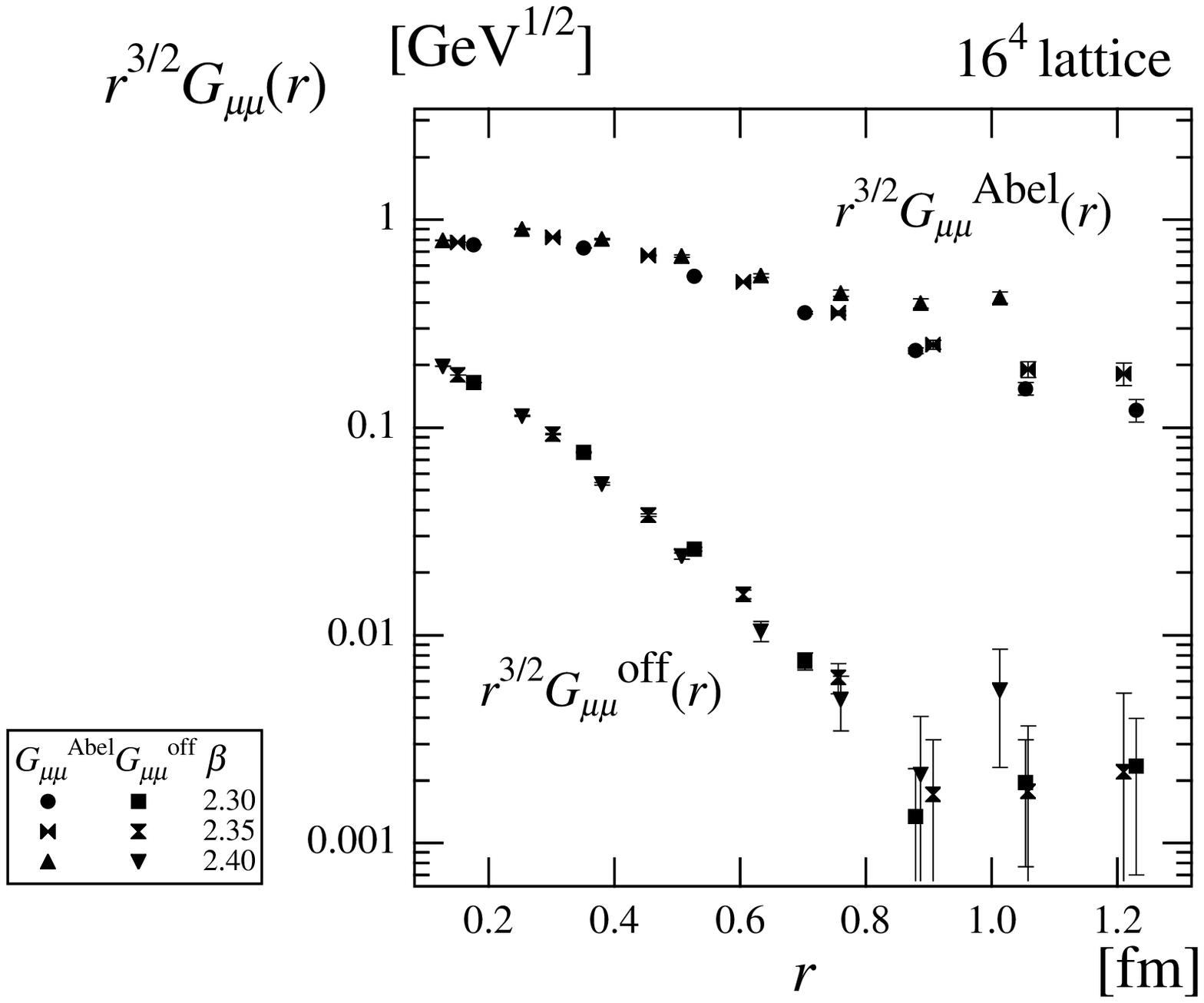} 
~\\
{\Huge Fig. 4(b)} 
\end{center}
\end{figure}
\begin{figure}
\begin{center}
\epsfig{figure=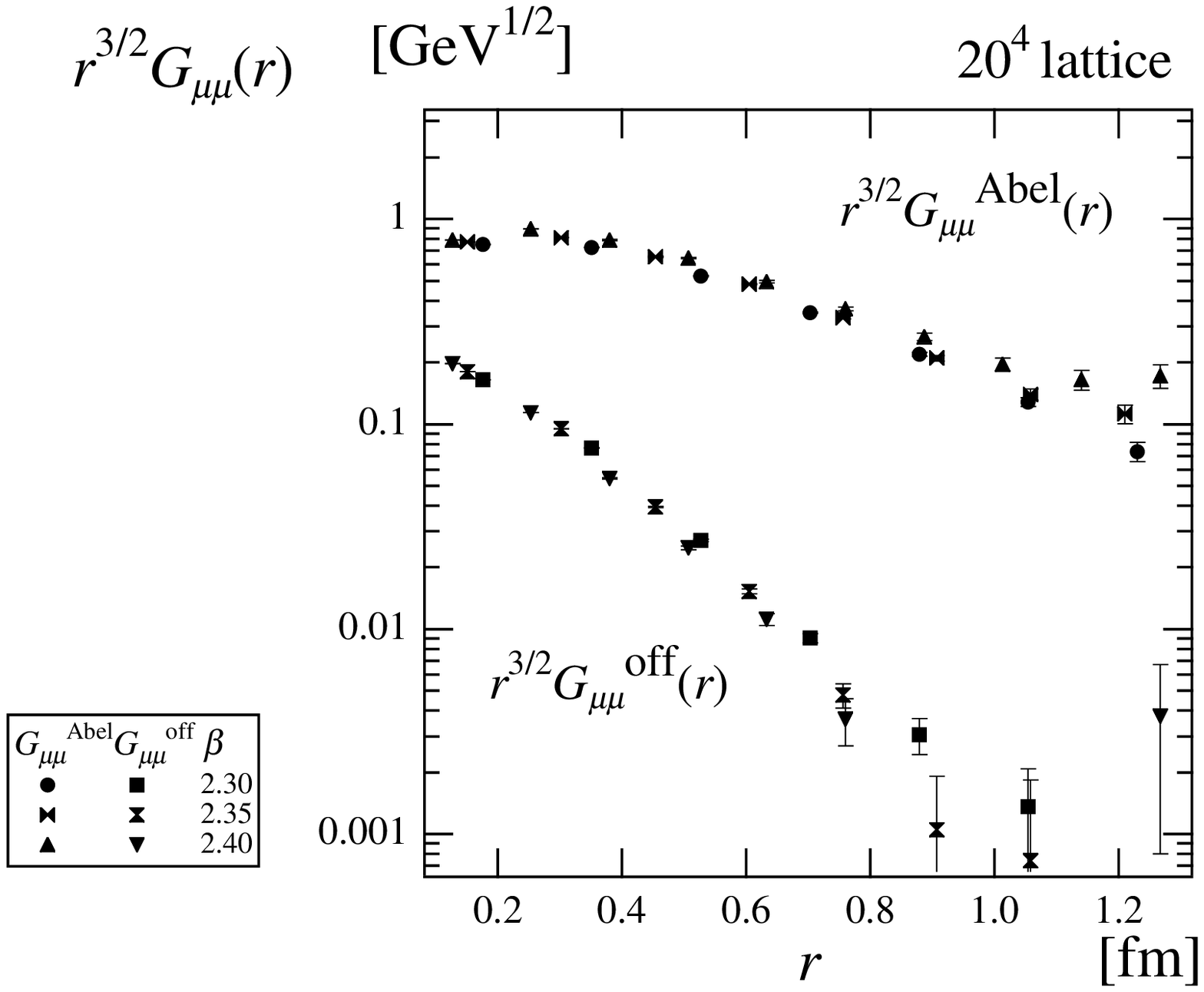} 
~\\
{\Huge Fig. 4(c)} 
\end{center}
\end{figure}
\begin{figure}
\begin{center}
\epsfig{figure=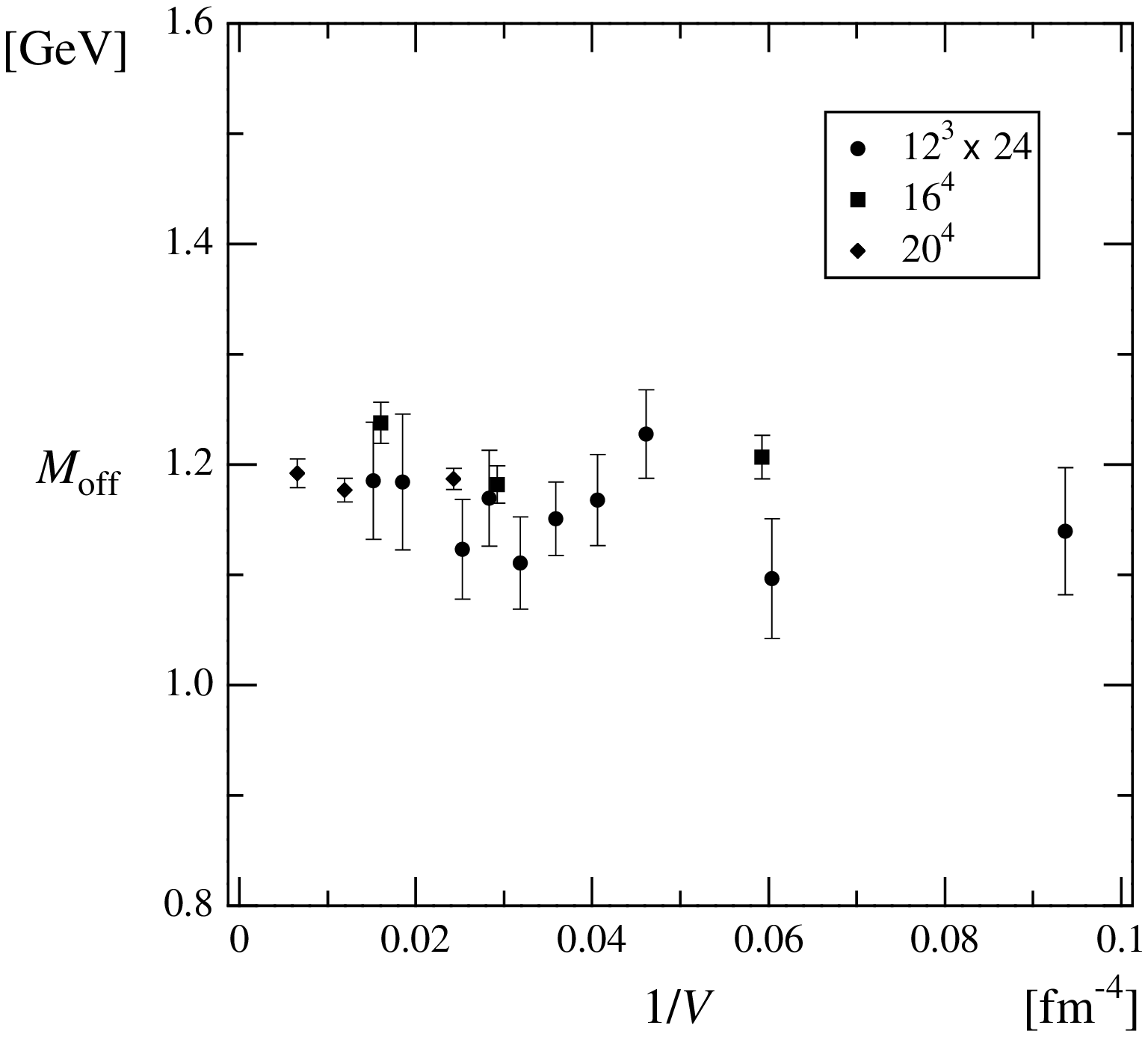} 
~\\
{\Huge Fig. 5} 
\end{center}
\end{figure}
\newpage
\label{caption:sigma}
\begin{center}
\begin{tabular}{ccc|lc}
\hline
$N_s$ & $N_\tau$ & $\beta$ & $\sqrt{\sigma}a$ & $a$ [fm]\\
\hline
\hline
8    &  10  & 2.20  & 0.4690(100) & 0.2233 \\
10   &  10  & 2.30  & 0.3690(30)  & 0.1757 \\
16   &  16  & 2.40  & 0.2660(20)  & 0.1267 \\
32   &  32  & 2.50  & 0.1905(8)   & 0.0907 \\
32   &  16  & 2.5115& 0.1836(13)  & 0.0874 \\
20   &  20  & 2.60  & 0.1360(40)  & 0.0648 \\
48   &  64  & 2.635 & 0.1208(1)   & 0.0575 \\
32   &  32  & 2.70  & 0.1015(10)  & 0.0483 \\
32   &  32  & 2.74  & 0.0911(2)   & 0.0434 \\
48   &  56  & 2.85  & 0.0630(30)  & 0.0300 \\
\hline
\end{tabular}
\end{center}
\begin{center}
{\Huge Table  1} 
\end{center}

\begin{center} 
\begin{tabular}{c|c|c|c}
\hline
  lattice size   & $\beta$     &   $\Meff $ & $\chi^2$ / $N_{df}$ \\
\hline
\hline
                 &    2.20     &  0.909(9)  & 21.8958 / 2 \\
                 &    2.23     &  0.921(9)  & 22.5831 / 2 \\
                 &    2.25     &  1.185(53) &  0.2457 / 2 \\
                 &    2.27     &  1.184(61) &  2.4423 / 2 \\
                 &    2.30     &  1.123(45) &  0.2110 / 2 \\
$12^3 \times 24$ &    2.31     &  1.170(44) &  0.5786 / 2 \\
                 &    2.32     &  1.111(42) &  0.5139 / 2 \\
                 &    2.33     &  1.151(33) &  0.9888 / 3 \\
                 &    2.34     &  1.168(41) &  0.6145 / 2 \\
                 &    2.35     &  1.228(40) &  4.0086 / 2 \\
                 &    2.37     &  1.096(54) &  9.0845 / 4 \\
                 &    2.40     &  1.140(58) & 11.9412 / 4 \\
\hline
                 &    2.30     &  1.238(19) &  4.0602 / 2 \\
$      16^4$     &    2.35     &  1.182(17) &  0.3777 / 3 \\
                 &    2.40     &  1.207(20) &  2.2900 / 4 \\
\hline
                 &    2.30     &  1.192(13) &  1.0963 / 2 \\
$      20^4$     &    2.35     &  1.177(11) & 10.7148 / 3 \\
                 &    2.40     &  1.187(10) &  6.4432 / 4 \\
\hline
\end{tabular}
\end{center} 
\begin{center}
{\Huge Table  2}
\end{center}


\newpage
\begin{thebibliography}{99}
\bibitem{GrRev} K. Sailer, Th. Sch$\ddot{\rm o}$nfeld, Zs. Schram, 
                A. Sch$\ddot{\rm a}$fer, and W. Greiner, 
		{ J. Phys. }{\bf B17}, 1005 (1991).
\bibitem{Rothe} H. J. Rothe, 
		\BOOK{Lattice Gauge Theories}
		{World Scientific, Singapore, 1992}.
\bibitem{Hay} 	Y. Peng and R. W. Haymaker, \Journal{\NPBPS}{34}{1996}{266};
		R. W. Haymaker, V. Singh, Y. Peng, and J. Wosiek, 
		\Journal{\PRD}{53}{1996}{389}.
\bibitem{YN}	Y. Nambu, \Journal{\PRD}{10}{1974}{4262}.
\bibitem{Drc} 	P. A. M. Dirac, 
		\Journal{ Proc. Roy. Soc. Lond. }{A133}{1931}{60}. 
\bibitem{Abr}	A. A. Abrikosov, { Sov. Phys. }JETP {\bf 5}, 1174 (1957).
\bibitem{GtH75}	G. 't Hooft, 
		\MEETtmp{in}
		{High Energy Physics}
		{edided by A. Zichichi}
		{Editorice Compositori, Bologna, 1975}.
\bibitem{SM}	S. Mandelstam,  \Journal{\PRC}{23}{1976}{245}.
\bibitem{GtH81}	G. 't Hooft,  \Journal{\NPB}{190}{1981}{455}.
\bibitem{SST}H. Suganuma, S. Sasaki, and H. Toki, 
		\Journal{\NPB}{435}{1995}{207}.
\bibitem{HS_SS_etal95}H. Suganuma, S. Sasaki, H. Toki, and H. Ichie, 
		{ Prog. Theor. Phys. Suppl.} 
		{\bf 120}, 57 (1995).
\bibitem{ichiemp} H.~Ichie and H.~Suganuma, preprint, 
		hep-lat/9808054.
\bibitem{umisedo} S.~Umisedo, H.~Suganuma, and H.~Toki, 
		\Journal{\PRD}{57}{1998}{1605}.
\bibitem{TS88and89}T.~Suzuki, {Prog.~Theor.~Phys.} {\bf 80}, 929 (1988);
		{\bf 81}, 752 (1989).
\bibitem{kondo}	K.-I.~Kondo, \Journal{\PRD}{57}{1998}{7467};
                \Journal{\PRD}{58}{1998}{105016}.
\bibitem{EI}	Z. F. Ezawa and A. Iwazaki, 
		{Phys. Rev.} D {\bf25}, 2681 {(1982)} ; 
		{\bf 26}, {631} {(1982)}.
\bibitem{diacomo} A.~Di~Giacomo,  \Journal{\NPBPS}{47}{1996}{136}
		and references therein.    
\bibitem{Sgnmykis}H. Suganuma, H. Ichie, A. Tanaka, and K. Amemiya, 
		\Journal{\PTPS}{131}{1998}{559}.
\bibitem{AT_inn}A.~Tanaka and H.~Suganuma, 
		\MEET{}
		{Proceedings of XVII RCNP International Symposimum 
		on Innovative Computational Methods 
		in Nuclear Many-Body Problems}
		{Osaka, Japan}
		{H. Horiuchi, M. Kamimura, H. Toki, Y. Fujiwara, M. Matuo, 
		and Y. Sakuragi}
		{World Scientific, Singapore, 1998}, p. 281 : 
		hep-lat/9712027.
\bibitem{SY}	T. Suzuki and I. Yotsuyanagi, \Journal{\PRD}{42}{1990}{4257}.
\bibitem{SH91}  S. Hioki, S. Kitahara, S. Kiura, Y. Matsubara, O. Miyamura, 
		S. Ohno, and T. Suzuki, \Journal{\PLB}{272}{1991}{326}.
\bibitem{ichiead} H.~Ichie and H.~Suganuma, 
               	\Journal{\NPB}{548}{1999}{365}.
\bibitem{OM}	O.~Miyamura, \Journal{\PLB}{353}{1995}{91}.
\bibitem{Wlf}	R.~M.~Woloshyn, \Journal{\PRD}{51}{1995}{6411}.
\bibitem{ASK}	A. S. Kronfeld, G. Schierholz, and U.-J. Wiese, 
		\Journal{\NPB}{293}{1987}{461}.
\bibitem{FB}	F. Brandstaeter, U.-J. Wiese, and G. Schierholz, 
		\Journal{\PLB}{272}{1991}{319}.
\bibitem{AS}	K. Amemiya and H. Suganuma, 
		\MEET{}
		{Proceedings of XVII RCNP International Symposimum 
		on Innovative Computational Methods 
		in Nuclear Many-Body Problems}
		{Osaka, Japan}
		{H. Horiuchi, M. Kamimura, H. Toki, Y. Fujiwara, M. Matuo, 
		and Y. Sakuragi}
		{World Scientific, Singapore, 1998}, p. 284 :
		hep-lat/9712028.
\bibitem{cheng}
		For instance, T.~P.~Cheng and L.~F.~Li,
		{\it Gauge Theory of Elementary Particle Physics}
		(Clarendon press, Oxford, 1984).
\bibitem{itzykson}For instance, C.~Itzykson and J.-B.~Zuber, 
		{\it Quantum Field Theory} (McGraw-Hill, New York, 1985).
\bibitem{MunstrLttcetxt} I. Montvay and G. M\"{u}nster,
		{\it Quantum Fields on a Lattice} 
		(Cambridge University press, Cambridge, 1994).
\bibitem{KA_HS_prep}K. Amemiya and H. Suganuma, in preparation.
\bibitem{HichHsugPRD} H.~Ichie and H.~Suganuma, 
	          \Journal\PRD{60}{1999}{077501}. %Phys. Rev. {D} in press.
\bibitem{HS_MF98}H. Suganuma, M. Fukushima, H. Ichie, and A. Tanaka,
		\Journal \NPBPS{65}{1998}{29}.
\bibitem{HI_etalPRD96}H. Ichie, H. Suganuma, and H. Toki,
		\Journal \PRD{54}{1996}{3382}.
\bibitem{HS_HI_etal_adelaid}H. Suganuma, H. Ichie, K. Amemiya, and A. Tanaka,
		\MEETtmp{in}
		{Proceedings of International Workshop 
		on Future Directions in Quark Nuclear Physics}
		{Adelaide, Australia, Mar. 1998, 
                 edited by A. W. Thomas and K. Tsushima}
		{World Scientific, Singapore, 1999}, p.44 : 
		hep-lat/9807034.
\bibitem{MO}	J. E. Mandula and M. Ogilvie \Journal{\PLB}{185}{1987}{127}.
\bibitem{ANaka}	A. Nakamura, H. Aiso, M. Fukuda, T. Iwamiya, T. Nakamura, and 
		M. Yoshida, 
		\MEET{}
		{Proceedings of the International Conference 
		on Color Confinement and Hadrons}
		{Osaka, Japan}
		{H. Toki, Y. Mizuno, H. Suganuma, T. Suzuki, and O. Miyamura}
		{World Scientific, Singapore, 1995},  p. 90;
		A. Nakamura, S. Sakai, and 
		Computational Science Division, National Aerospace Laboratory,
		\Journal{\PTPS}{131}{1998}{585}.
\bibitem{MO90}  J. E. Mandula and M. Ogilvie, \Journal{\PLB}{248}{1990}{156}. 
\bibitem{CMicMTep} C. Michael and M. Teper, \Journal{\PLB}{199}{1987}{95}.
\bibitem{UKQCD93} UKQCD Collaboration, \Journal{\NPB}{394}{1993}{509}.
\bibitem{CMicSPer} C. Michael and S. Perantonis, 
                \Journal{\NPBPS}{20}{1991}{177}.
\bibitem{JFinUHelFKar}  J. Fingberg, U. Heller and F. Karsch, 
                \Journal{\NPB}{392}{1993}{493}.
\bibitem{BalCSchKSch} G. S. Bali, C. Schlichter and K. Schilling, 
                \Journal{\PRD}{51}{1995}{5165}.
\bibitem{Huang} For instance, K.~Huang,
		{\it Quarks, Leptons and Gauge Fields}  
		(World Scientific, Singapore, 1992).
\bibitem{BaSchlSchi}
                G. S. Bali, V. Bornyakov, M. M\"{u}ller-Preussker, and K. Schilling, 
                \Journal{\PRD}{54}{1996}{2863}.
\bibitem{BorMitMulPahl} 
                V. G. Bornyakov, V. K. Mitrjushkin, M. M\"{u}ller-Preussker, and F. Pahl, 
                \Journal{\PLB}{317}{1993}{596}.
\end{thebibliography}
\end{document}